\documentclass[
     twocolumn,
 	prd,
 	amssymb,
 preprintnumbers,superscriptaddress,
 	nofootinbib]{revtex4-2}
  
\usepackage{amsmath,amsfonts,amssymb,enumitem}
\usepackage{mathrsfs}
\usepackage{graphicx}
\usepackage[english]{babel} 
\usepackage{url} 
\usepackage{bbold}
\usepackage[utf8]{inputenc} 
\usepackage[colorlinks=true,citecolor=blue,urlcolor=blue]{hyperref}
\usepackage{multirow} 
\usepackage{booktabs}
\pdfoutput=1
\usepackage{paralist}
\usepackage{graphicx}
\usepackage{latexsym}
\usepackage{xcolor}
\usepackage[export]{adjustbox}
\usepackage[thinlines]{easytable}
\usepackage{slashed}
\usepackage{dcolumn}
\usepackage{verbatim}
\usepackage{float}
\usepackage{multirow}
\usepackage{xspace}
\usepackage[normalem]{ulem}
\usepackage{tabularx}

\newcommand{\LCDM}{$\Lambda$CDM}
\newcommand{\wowaCDM}{$w_0 w_a$CDM}

\newcommand{\nodata}{$\cdots$}

\begin{document}

\title{First Constraints on a Pixelated Universe in Light of DESI}

\author{Jonathan J.~Heckman}
\email{jheckman@sas.upenn.edu}

\affiliation{Department of Physics and Astronomy, University of Pennsylvania, Philadelphia, PA 19104, USA}

\author{Omar F.~Ramadan}
\email{oramadan@hawaii.edu}

\author{Jeremy Sakstein}
\email{sakstein@hawaii.edu}
\affiliation{
Department of Physics $\&$ Astronomy, University of Hawai‘i,
Watanabe Hall, 2505 Correa Road, Honolulu, HI, 96822, USA}

\date{June 2024}

\begin{abstract}
Pixelated dark energy is a string theory scenario with a quantum mechanically stable cosmological constant.~The number of pixels that make up the  universe slowly increases, manifesting as a time-dependent source of dark energy.~DESI has recently reported evidence for dynamical dark energy that fits within this framework.~In light of this, we perform the first cosmological analysis of the pixelated model.~We find that the simplest model where the pixel growth rate is constant is able to accommodate the data, providing a marginally better fit than $\Lambda$CDM;~and we show that models where the pixel growth rate is increasing and of order the Hubble constant today could provide better fits.~Our analysis helps to clarify the features of UV constructions of dark energy necessary to accommodate the data.
\end{abstract}
\maketitle

\section{Introduction}

The microscopic origin of dark energy remains one of the outstanding issues at the interface of high energy physics and cosmology.~While \LCDM\ provides an economical framework which accommodates most observational constraints \cite{Planck:2018vyg}, the construction of UV complete models remains an important topic of current research.\footnote{For recent reviews of the rather large literature, see e.g., \cite{Cicoli:2023opf,VanRiet:2023pnx} and references therein.}~In addition, recent results from DESI \cite{DESI:2024mwx}  indicate a preference for a dynamic component of dark energy, which has stimulated multiple investigations interpreting the data within the context of competing dark energy models \cite{Colgain:2024xqj,Ramadan:2024kmn,Bhattacharya:2024hep,Carloni:2024zpl,Park:2024jns,Cortes:2024lgw,Shlivko:2024llw,Dinda:2024kjf,Wang:2024hks,Luongo:2024fww,Wang:2024dka,Tada:2024znt,Yin:2024hba,Berghaus:2024kra,Andriot:2024jsh,Yang:2024kdo}.~Given this state of affairs, it is natural to seek out UV-motivated dark energy scenarios and investigate which, if any, are favored by data. 

In this note we revisit the pixelated dark energy scenario\footnote{We refer to this as a scenario since a complete top-down implementation of the proposal remains to be carried out.} of references \cite{Heckman:2018mxl, Heckman:2019dsj} which is motivated by top-down string / F-theory considerations.~The main idea in this proposal is that at low energies F-theory on a $\mathrm{Spin}(7)$ background leads to a 4D model which in Kleinian, i.e., $2+2$ spacetime signature preserves $\mathcal{N} = 1/2$ supersymmetry.~In Lorentzian, i.e., $3+1$ signature this means there is no supersymmetry, but the faint remnant of supersymmetry in Kleinian signature yields states which remain stable against radiative corrections \cite{Heckman:2018mxl, Heckman:2019dsj, Heckman:2022peq}.~For additional details on the underlying UV details of this model, as well as the constraints imposed by the DESI results, see Appendix \ref{app:UV}.

Our interest here is in the potential observational consequences of this proposal in cosmology.~The main elements consist of an (unstable) Einstein static Universe of topology $\mathbb{R} \times \mathrm{S}^3$ with a balancing between a stiff fluid $(w = +1)$ and dark energy $(w = -1)$ which are respectively generated by a three-form flux threading the $S^3$ and an internal gradient of a dilaton.\footnote{We note that, while the model has positive spatial curvature, the requirements that inflation is reproduced implies that $\Omega_\kappa$ is far smaller than the observational bounds \cite{Heckman:2019dsj} so for all intents and purposes the universe is flat.~We will therefore neglect spatial curvature in what follows.}~The presence of fluxes can be viewed as a collection of branes tiling the $S^3$, i.e., ``pixels'' each of which contributes one unit of flux.~We refer to the total number of pixels as $N$.

{In the extra-dimensional geometry, each pixel is associated with a five-brane which wraps a locally stable five-cycle in the internal geometry \cite{Heckman:2018mxl}.~Globally, however, this five-cycle is topologically trivial and as such, the number of such branes can either ``wind up'' leading to an increase in the number of pixels or ``unwind'' leading to a decrease in the number of pixels.~Consequentially, the total number of pixels is a function of time $N(t)$.~When $N$ is large, this can be approximated as a continuous function, but it is worth noting that there is always some intrinsic discretization in this parameter.~In \cite{Heckman:2019dsj} a model of pixel growth based on the emission of radiation from pixel production was introduced.~In this model, the initial presence of radiation stimulates the production of pixels in the 4D Universe.~This stimulated emission of pixels then leads to the emission of additional radiation, setting off a chain reaction with a constant growth rate in the number of pixels.~There can be deviations from this constant growth rate due to quantum tunneling events in the pixel / flux number, but clearly the simplest and best motivated scenario is one in which this growth rate $\Gamma \equiv \dot{N} / N$ is a constant.~In principle other UV motivated scenarios could accommodate a time dependent growth rate, and taking the DESI results at face value, we will indeed show that, while the case of constant $\Gamma$ can accommodate the data, a time-varying growth rate could provide a better fit.~This in turn motivates the investigation of a broader class of UV motivated scenarios with accelerated pixel growth.

To match these UV considerations to cosmology, we need to explain how this scenario predicts a time-varying equation of state for dark energy. The unstable nature of the solution leads to a  time dependence on the number of pixels, and, consequentially, a time-dependent cosmological ``constant'' 
\begin{equation}
\label{eq:CC}
\Lambda(t) = \frac{8 \pi^2}{\ell_s^2} \frac{1}{N(t)},
\end{equation}
where $\ell_s$ denotes the string length and the number of pixels $N(t)$ is time-dependent.~From this one 
obtains a time-dependent equation of state (EOS) for dark energy $P = w \rho$ with \cite{Heckman:2019dsj}:
\begin{equation}
\label{eq:wFull}
w(t) = -1 + \frac{\dot{N}(t)}{3H(t)N(t)}.
\end{equation}
{It is important to note that $w(t)$ as derived above is not $P/\rho$ for some fundamental perfect fluid e.g., a quintessence scalar but, rather, it is what is inferred if one treats the time-dependent cosmological constant in equation \eqref{eq:CC} as an effective fluid at the level of the Friedmann and continuity equations and reverse-engineers what $w(t)$ must be to ensure consistency of the system.~Importantly, $w(t)$ in Eq.~\eqref{eq:wFull} was not found by integrating the equation of motion for new degrees of freedom.~Determining its full form must be accomplished on a model-by-model basis in the  string theory construction to derive $N(t)$.~The simplest scenario of constant pixel growth rate will be shown to be marginally preferred over $\Lambda$CDM.~Beyond this, there is no canonical model for $N(t)$ but observational progress can be made in a model-independent manner by parameterizing $N(t)$ in terms of its derivatives at the current time and constraining these with low-redshift probes.~The resulting bounds then clarify the features of the UV construction necessary to accommodate the data as discussed further in Appendix~\ref{app:UV}}.

{The Chevallier-Polarski-Linder (CPL) or $w_0$--$w_a$ parameterization of dynamical dark energy \cite{Chevallier:2000qy,Linder:2002et} 
\begin{equation}
\label{eqn:w(a)}
    w(a)  = w_0 + w_a(1-a) 
\end{equation}
provides a convenient and well-studied framework for characterizing the cosmological effects of $N(t)$.~Expanding Eq.~\eqref{eq:wFull} around $a = 1$, one finds that the equation of state fits within this framework with \cite{Heckman:2019dsj}}
\begin{align}
w_0  & = -1 + \frac{\dot{N}}{3 H_{0} N} \label{eqn:w0} \textrm{ and}\\ 
w_a  & = - \frac{1}{2} \Omega_{m,0} \frac{\dot{N}}{H_{0} N} .~\label{eqn:wa}
\end{align}
In the regime of interest, i.e., close to present day values, $\dot{N} > 0$ so $w_0 > -1$ and $w_a < 0$.~The quantity $\dot{N}/N=\Gamma$ is the growth rate/decay rate of the pixels (depending on the sign of the derivative).~In deriving equations~\eqref{eqn:w0} and \eqref{eqn:wa} it was assumed that $\Gamma$ is approximately constant i.e., $\dot{\Gamma}/{\Gamma}\ll H$ and that $\Gamma/H_0\ll1$ so that higher-order terms may be neglected \cite{Heckman:2019dsj}.~While this is the simplest choice, one can in principle consider a time-dependent $\Gamma$. We revisit this possibility later.~We derive equations~\eqref{eqn:w(a)}--\eqref{eqn:wa} in Appendix \ref{sec:derivation} where we also use our results to confirm that when $\Gamma$ is assumed to be constant the approximation $\Gamma/H_0\ll1$ is valid \textit{a posteriori}.~

{While Pixelated DE fits within the broader CPL parameterization, it is more restrictive.~The CPL parameterization is a two-parameter extension of $\Lambda$CDM because $w_0$ and $w_a$ are independent parameters but, in contrast, Pixelated DE (with constant $\Gamma$) is a one-parameter extension because equations~\eqref{eqn:w0} and \eqref{eqn:wa} imply the relation
\begin{equation}
    \label{eq:oneParama}1+w_0=-\frac{2}{3\Omega_{m,0}}w_a.
\end{equation} 
In addition, the appearance of $\Omega_{m,0}$ in this relation is also distinct from CPL, which imposes that $w_0$ and $w_a$ are independent of $\Omega_{m,0}$.~This parameter is jointly constrained in cosmological data analyses so the DESI bounds reported in \cite{DESI:2024mwx} cannot be mapped directly onto Pixelated DE because these analyses do not account for possible covariances induced by the relation in Eq.~\eqref{eq:oneParama}.~Indeed, reference \cite{Colgain:2024xqj} noted that the DESI covariances of  $\Omega_{m,0}$ with other cosmological parameters is large.~It is therefore essential that the relation in Eq.~\eqref{eq:oneParama} is fit to DESI, or,  more generally, to the full combination of cosmological data sets considered, in an independent analysis.~Motivated by this, we now test the pixelated DE model against cosmological observations for the first time.}

\begin{table*}[ht]
\centering
\setlength{\tabcolsep}{10pt}
\def\arraystretch{1.5}
\resizebox{\textwidth}{!}{%
    \small
    \begin{tabular}{lccc}
    \toprule
    \midrule
    Parameter \text{\&} Model & {\textbf{Flat}} $\boldsymbol{\Lambda}${\textbf{CDM}} & $\boldsymbol{w_a w_a}${\bf CDM} & \textbf{Pixelated Dark Energy} \\
    \midrule
    \textbf{Sampled Parameters} &&&\\
    $\log(10^{10} A_\mathrm{s})$  & $3.051(3.048)\pm 0.013$ & $3.041(3.041)\pm 0.013$ & $3.058(3.058)\pm 0.014$ \\ 
    $n_\mathrm{s}$ & $0.9674(0.9664)\pm 0.0035$ & $0.9658(0.9676)\pm 0.0038$ & $0.9694(0.9697)\pm 0.0038$ \\ 
    $\Omega_b h^2$ & $0.02243(0.02243)\pm 0.00013$ & $0.02238(0.02237)\pm 0.00014$ & $0.02249(0.02251)\pm 0.00013$ \\
    $\Omega_{c} h^2$ & $0.11907(0.11931)\pm 0.00083$ & $0.11963(0.11948)\pm 0.00097$ & $0.11835(0.11858)\pm 0.00091$ \\
    $100\theta_*$ & $1.04195(1.04207)\pm 0.00028$ & $1.04189(1.04184)\pm 0.00029$ & $1.04203(1.04198)\pm 0.00028$ \\
    $\tau_\mathrm{reio}$ & $0.0579(0.0569)\pm 0.0069$ & $0.0530(0.05337)\pm 0.0073$ & $0.0615(0.0613)^{+0.0068}_{-0.0088}$ \\
    $w_0$ & $...$ & $-0.733(-0.705)\pm 0.068$ & $-0.949(-0.955)\pm 0.025$ \\
    $w_a$ & $...$ & $-1.01(-1.01)^{+0.34}_{-0.28}$ & $-0.025(-0.021)^{+0.015}_{-0.013}$ \\
    \midrule
    \textbf{Derived Parameters} &&&\\
    $H_0 \left[\text{km/s/Mpc}\right]$ & $67.78(67.73)\pm 0.37$ & $67.21(67.13)\pm 0.65$ & $66.76(66.85)^{+0.66}_{-0.59}$ \\
    $\Omega_m $ & $0.3095(0.3104)\pm 0.0049$ & $0.3159(0.3163)\pm 0.0065$ & $0.3175(0.3172)\pm 0.0064$ \\
    \midrule
    $\boldsymbol{\chi^2}$\textbf{ Statistics} &&&\\
    $\chi^2_{\rm bf}(\Delta)$ & $4450.75$ & $4434.11(-16.64)$ & $4446.99(-3.76)$ \\ 
    $\chi^2_{\rm bf, CMB}(\Delta)$ &  $2771.89$ &  $2770.62(-1.27)$ &  $2773.56(+1.67)$ \\ 
    $\chi^2_{\rm bf, BAO}(\Delta)$ & $17.82$ & $11.93(-5.89)$ & $18.50(+0.68)$ \\
    $\chi^2_{\rm bf, SNe}(\Delta)$ & $1646.51$ & $1637.82(-8.69)$ & $1640.80(-5.71)$ \\ 
    $\chi^2_{\rm bf}/{\rm DoF}$ & $1.072$ & $1.068$ & $1.071$ \\ 
    Tension Level & \nodata & $3.49\sigma$ & $1.62\sigma$\\
    \midrule
    \bottomrule
    \end{tabular}
}
\caption{
    Marginalized posteriors for flat \LCDM, $w_0w_a$CDM, and pixelated DE models using CMB+DESI+DESY5 likelihoods, showing the mean(best fit) and the $68\%$ confidence interval.~The \LCDM\ parameters share the same prior across models and only the priors of $\{w_0, w_a\}$ parameters differ in $w_0 w_a$CDM and pixelated DE.~We also show the best fitting $\chi^2_{bf}(\Delta)$, where $\Delta=\chi^2_{\rm bf,model} - \chi^2_{\rm bf,\Lambda\text{CDM}}$ is the difference between the best fitting $\chi^2$ values, subtracted from that of \LCDM.~The statistically significant tension levels with \LCDM\ are reported as well.~\label{tab:marg_param}
}
\end{table*}

\section{Method and Results}

We modified the cosmological solver \texttt{CLASS} \cite{Blas:2011rf,Lesgourgues:2011re} to evolve the cosmology of pixelated DE according to equation \eqref{eqn:w(a)} with the relation $\eqref{eq:oneParama}$.~For the standard $w_0$--$w_a$ parameterization, we sampled over the 6 \LCDM\ parameters $\{A_s, n_s, 100\theta_*, \Omega_b h^2, \Omega_{c} h^2, \tau_{reio}\}$ plus $\{w_0, w_a\}$.~However, for pixelated DE, equation~\eqref{eq:oneParama} implies that $w_a$ is a derived parameter given by a combination of the sampled $\Omega_{m,0}$ and $w_0$, so we only sampled over $\{A_s, n_s, 100\theta_*, \Omega_b h^2, \Omega_{c} h^2, \tau_{reio}, w_0\}$.~We fitted both models to the combination of cosmic microwave background (CMB) data, specifically the Planck 2018 CMB spectra \cite{Planck:2018vyg}, CMB gravitational lensing from a combination of Planck 2020 lensing \cite{Planck:2020olo,Carron:2022eyg} and ACT DR6 \cite{ACT:2023dou,ACT:2023kun};~the DESI BAO measurements \cite{DESI:2024mwx};~and the DESY5 \cite{Abbott:2024agi} supernovae compilation.~This combination of data is identical to that used by the DESI analysis \cite{DES:2024tys}.~The Markov Chain Monte Carlo (MCMC) sampling was performed using the  \texttt{Cobaya} code \cite{Torrado:2020dgo};~the sampler was deemed to have converged when the standard Gelman-Rubin criteria $R-1<0.01$ \cite{GelmanRubin1992} was achieved.~For  pixelated DE, we imposed the priors $w_0 \sim \mathcal{U}[-1, 1]$, which reflect the requirement in Eq.~\eqref{eqn:w(a)} that $w(a)\ge-1$ in pixelated DE, while for the $w_0$-$w_a$ parameterization we imposed a wide prior with $w_0 \sim \mathcal{U}[-3, 1]$ and $w_a \sim \mathcal{U}[-3, 2]$.~We analyzed and plotted our chains using \texttt{GetDist} \cite{Lewis:2019xzd}.~For comparative purposes, we also fitted a general CPL $w_0$--$w_a$ model.~Our results are given in table~\ref{tab:marg_param}, with 2D contour plots and 1D marginalized posteriors given in figure \ref{fig:full_contours}.

For the general CPL model, we find consistent results with DESI \cite{DESI:2024mwx}:~$w_0=-0.733\pm0.068$, $w_a=-1.01^{+0.34}_{-0.28}$ and confirm their findings that this model is preferred over $\Lambda$CDM at the $\sim3.5\sigma$ level.~The pixelated DE relation yielded $w_0=-0.949\pm0.025$ and $w_a=-0.024^{+0.015}_{-0.013}$.~We derived the pixel growth rate with its $68\%$ confidence interval:
\begin{equation}
    \frac{\Gamma_0}{H_0}=0.145\pm 0.075
\end{equation}
with the best fitting value $(\Gamma_0/H_0)_{\textrm{best fit}} = 0.135$, indicating that the number of pixels is increasing at a smaller rate than the expansion of the universe.

\begin{figure}[t]
\centering
\includegraphics[width=3.33in]{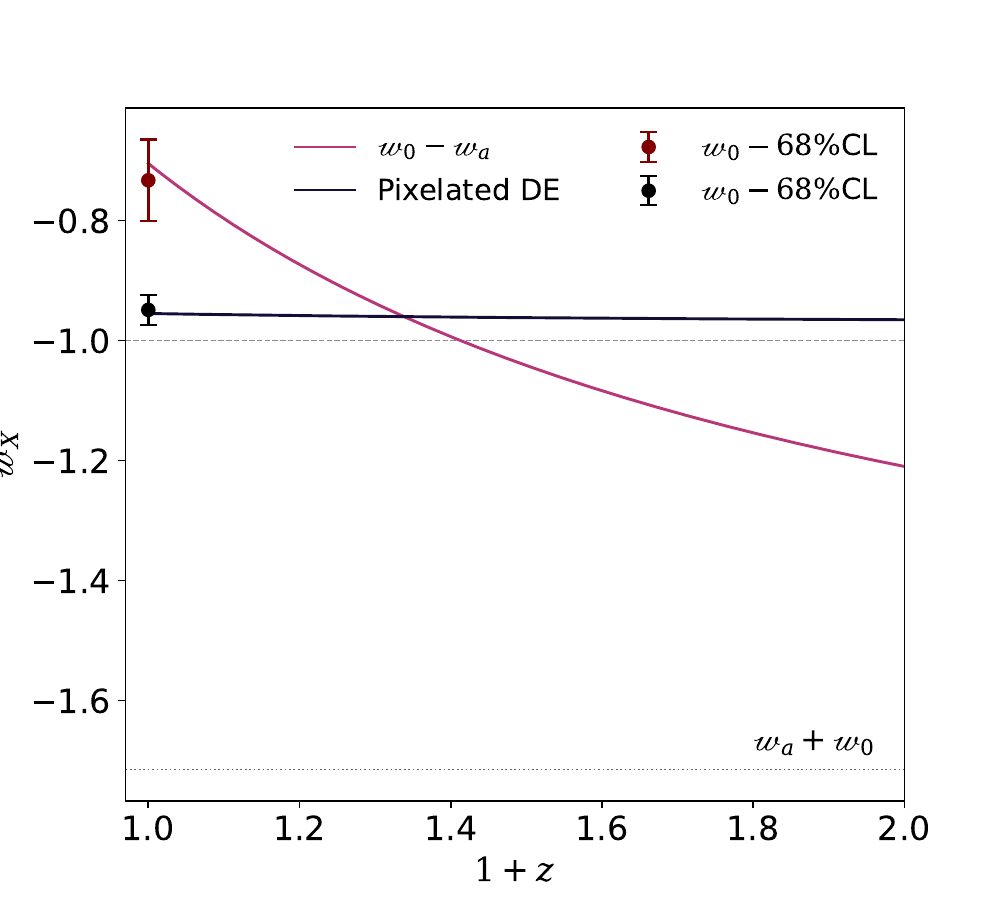}
    \caption{The best fitting EOS for the \wowaCDM\ model and pixelated dark energy.~The CPL parameterization suggests that the dark energy fluid initially has an equation of state of $w_a + w_0=-1.72$, represented by the dotted line, for $z\gtrsim10^3$.~The EOS evolves around redshift $z\approx10^3$, crossing the phantom line $w=-1$ at $z\approx0.4$, coinciding with the onset of dark energy domination.~The black solid line is the EoS of pixels in which the field starts frozen at $w=-0.976$ and evolves to $w_0 = -0.955$ today.~The dashed line represents the EOS for a cosmological constant, $w_\Lambda = -1$.~We also show the $68\%$ confidence level posteriors of $w_0$ for both $w_0 w_a$CDM and pixelated DE in red and black, respectively.
    }
\label{fig:EoS_w0wa_bestfit}
\end{figure}

The data showed a mild preference for the Pixelated DE model over \LCDM~at the level of $1.6\sigma$ with an improvement in the best-fit by an amount $\Delta\chi^2_{\textrm{best fit}}=-3.76$ for one additional degree of freedom.~The  improvement suggests that the additional complexity of our model improves the fit compared to the simpler \LCDM, but not as well the  more generalized CPL model, which is preferred at $3.5\sigma$ over \LCDM~with $\Delta\chi^2_{\textrm{best fit}}=-16.64$.~From table~\ref{tab:marg_param}, we see that the preference for Pixelated DE over \LCDM~is driven by the fit to the DESY5 supernovae dataset, which alone improves the fit by $\Delta\chi^2_{\textrm{bestfit}} = -5.71$ compared to \LCDM. However, the Pixelated DE model underperforms relative to \LCDM~when considering the remaining datasets combined.~The reason for this is that the relation between $w_0$ and $w_a$ in equation (\ref{eq:oneParama}) does not allow $w(z)$ to vary rapidly enough to provide a close fit to the data at all redshifts.~To elaborate, the supernovae and DESI data at redshift $z\lesssim0.51$ deviates from the $\Lambda$CDM prediction while higher redshift  data is consistent (see \cite{DESI:2024mwx,Colgain:2024xqj, Shlivko:2024llw,Ramadan:2024kmn, Bhattacharya:2024hep,Abbott:2024agi} for additional discussion on this).~The CPL model is able to fit the data at all redshifts by having $w_0>-1$ and $w_a$ negative and large in magnitude, with $w_a\sim2w_0$  so that $w(z)$ approaches $-1$ rapidly when $z>0.51$.~On the other hand, the $w_0$--$w_a$ model corresponding to pixelated DE does not allow for this.~As demonstrated by equations~\eqref{eqn:w0} and~\eqref{eqn:wa}, $w_0$ differs from $-1$ by an amount of order $\Gamma/H_0$ so fitting the low-$z$ data requires  $\Gamma/H_0\sim\mathcal{O}(10^{-1})$ implying that $w_a\sim\mathcal{O}(10^{-2})$  because it is suppressed by a factor of $\Omega_{m,0}\sim0.3$ relative to $\Gamma/H_0$.~Thus, the EOS evolves too slowly over the range of relevant redshifts to fully accommodate each data point.~This is exemplified in figure~\ref{fig:EoS_w0wa_bestfit} where we plot the best fitting CPL and pixelated DE equation of state as a function of redshift.~At $z>0$ the CPL model EOS rapidly moves from $w>-1$ towards $w=-1$ but the pixelated DE EOS barely evolves.

\section{Outlook}

The considerations above provide guidance for constructing UV complete models of dark energy that are able to better-fit the data.~Clearly, breaking the relation between $w_0$ and $w_a$ that forces $|w_a|<w_0$ is necessary.~This can be accomplished by moving beyond the approximation that the pixel growth rate $\Gamma$ is  constant.~We derive the CPL parameterization for this more general case in Appendix~\ref{sec:derivation} for the first time.~We find
\begin{align}
    \label{eq:w0genGamma}
    w_0&=-1+\frac{\Gamma_0}{3H_0}\textrm{ and}\\
\label{eq:waGenGamma}
w_a&=- \frac{1}{2} \Omega_{m,0}\frac{\Gamma_0}{H_0}-\frac{\Gamma_1}{3H_0^2}-(1-\Omega_{m,0})\frac{\Gamma_0^2}{6H_0^2},
\end{align}
where:
\begin{equation}
\label{eq:Gamma1DefSpec}
\Gamma_0=\left.\frac{\dot{N}}{N}\right\vert_0\quad\textrm{ and }\quad \Gamma_1 = 
\left.\frac{\ddot{N}}{N^2}\right\vert_0 - \Gamma_0^2,
\end{equation}
with subscript zeros indicating quantities evaluated at the present time $t_0$.~Thus, in the general case, the equation of state also depends on $\ddot{N} / N^2$.~

This gives a less restricted two-parameter model that can be fit to the data to bound $\Gamma_0$ and $\Gamma_1$.~The CPL bounds cannot be directly translated into bounds on $\Gamma_0$ and $\Gamma_1$ because, in the pixelated model, $w_a$ depends on $\Omega_{m,0}$, which is simultaneously varied in the MCMC so a full fit to the data is required to account for this.~We can however estimate $\{\Gamma_0,\,\Gamma_1\}$ by using the best-fitting parameters in our \wowaCDM\ analysis.~Doing so, we estimated the $68\%$ confidence interval
\begin{equation}
\label{eq:Omar'sGamma}
    \frac{\Gamma_0}{H_0}=0.8\pm 0.20\quad \textrm{ and }\quad \frac{\Gamma_1}{H_0^2}=2.41\pm 0.72,
\end{equation}
with the best fitting values $(\Gamma_0/H_0)_{\textrm{best fit}}=0.88$ and $(\Gamma_1/H_0^2)_{\textrm{best fit}}=2.61$.~In terms of the pixel number acceleration, we found the $68\%$ CL  
\begin{equation}
    \frac{\ddot{N}}{N^2 H_0^2}=3.09^{+0.93}_{-1.2} 
\end{equation}
with best fitting value $[\ddot{N}/(NH_0)^2]_{\textrm{best fit}} = 3.39$.~Thus, the data indicate that the pixel growth rate today is comparable with the expansion rate of the universe, and is beginning to increase.~Deriving the full time-dependence of the pixel decay rate in the UV-construction and determining whether this scenario can be realized is then paramount.\footnote{As an example, \cite{Heckman:2019dsj} also found there can be larger jumps in the number of pixels from thermal and quantum events, leading to an acceleration in the pixel number.} Turning the discussion around, a detailed fit with the extra parameter $\Gamma_1$ would provide important hints on the microphysical dynamics of pixels (see Appendix \ref{app:UV}).

\begin{figure*}
\centering
\includegraphics[width=7.05in]{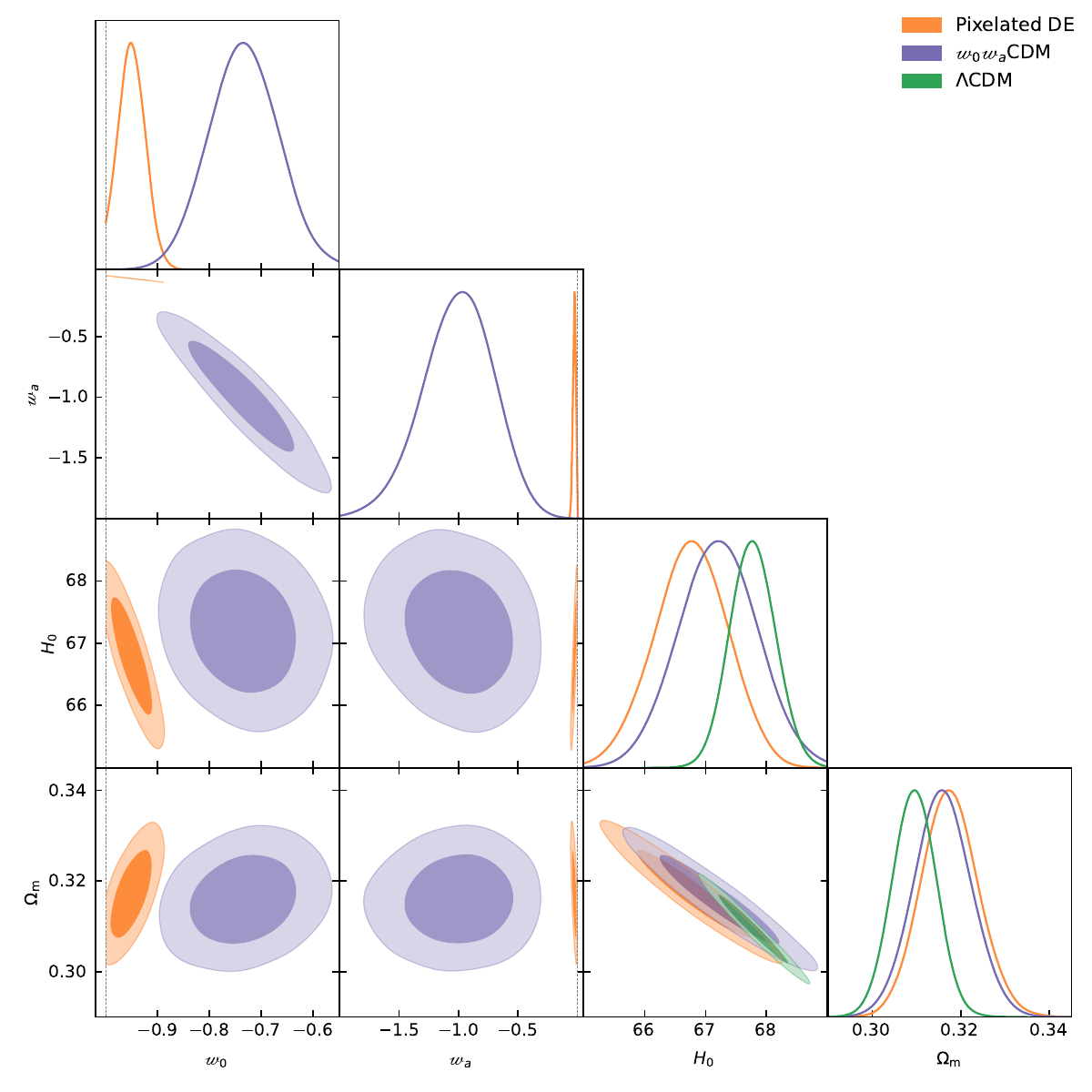}
    \caption{Marginalized posteriors for different cosmologies fitted to CMB+DESI+DESY5.~The inner contours represent the $68\%$ confidence level (CL) where the outer is $95\%$ CL.~The dashed lines indicate the \LCDM\ limit with $w_0=-1$ and $w_a = 0$.~Both $w_0 w_a$CDM and pixelated DE include the \LCDM\ limit, with pixelated DE being marginally preferred over \LCDM.}\label{fig:full_contours}
\end{figure*}

\begin{acknowledgments}
We are grateful for discussions with Jason Kumar, Craig Lawrie, David Rubin, Greg Tarl\'{e}, and Xerxes Tata.~The technical support and advanced computing resources from University of Hawai‘i Information Technology Services – Cyberinfrastructure, funded in part by the National Science Foundation CC\* awards \#2201428 and \#2232862 are gratefully acknowledged.~The work of JJH is supported by DOE (HEP) Award DE-SC0013528.
\end{acknowledgments}

\appendix

\section{CPL Parameterization For Pixelated Dark Energy}
\label{sec:derivation}

In this Appendix, we derive equations~\eqref{eq:w0genGamma} and \eqref{eq:waGenGamma}, which have not previously appeared in the literature.~Our starting point is the EOS for pixelated DE derived by \cite{Heckman:2019dsj} given in equation~\eqref{eq:wFull}.~The late-time universe is well-approximated as consisting of matter and a time-dependent cosmological constant given by $\Lambda(t)=\Lambda_0/N(t)$ with $\Lambda_0=8\pi^2/l_s^2$ (see equation \eqref{eq:CC}) so the Friedmann equation is
\begin{equation}
\label{eq:PixelatedFriedmann}
H^2(t)=\frac{\Omega_{m,0}H_0^2}{a(t)^3}+\frac{H_0^2(1-\Omega_{m,0})N_0}{N(t)},
\end{equation}
where $N_0=N(t_0)$ is the number of pixels at the present time, and $\Lambda_0$ has been replaced by $N_0$ using the relation $\Lambda_0=3H_0^2N_0(1-\Omega_{m,0})$.~One can then Taylor expand Eq.~\eqref{eq:PixelatedFriedmann} to first-order in $(1-a)$, similarly expand $\Gamma(t)$, and substitute both into equation~\eqref{eq:wFull} to find the CPL parameterization with $w_0$ and $w_a$ given in equations~\eqref{eq:w0genGamma} and~\eqref{eq:waGenGamma} with $\{\Gamma_0,\,\Gamma_1\}$ given in Eq.~\eqref{eq:Gamma1DefSpec}.

In the constant $\Gamma_0$ analysis above, we made the approximation that $\Gamma_0/H_0\ll1$.~We now verify that this was justified.~The best fitting model gave $\Gamma_0/H_0=0.078$ so the error in $w_a$ by neglecting the final two terms in Eq.~\eqref{eq:waGenGamma}  is $\Delta w_a=2.7\times10^{-3}$, an order of magnitude smaller than both the best fitting value of $w_a$ and its error bars.

\section{UV Origin of Pixelated Dark Energy} 
\label{app:UV}

In this Appendix we provide further details on the UV motivation for the pixelated dark energy scenario. We refer to this as a scenario because while many of the ingredients are quite plausible in the context of a string compactification, a full realization in terms of an explicit compactification geometry remains an outstanding open problem. Rather, we view the present work as imposing observational constraints on this class of scenarios.

To set the stage, we first recall how 4D $\mathcal{N} = 1$ theories arise in the context of F-theory compactifications. We refer the interested reader to 
\cite{Heckman:2010bq, Weigand:2018rez} for additional details, as well as 
references to the primary literature. In F-theory realizations of 4D Minkowski backgrounds, one considers an elliptically fibered Calabi-Yau fourfold. In this setting, the elliptic fibration serves as a proxy for determining the location of 7-branes wrapped over divisors of a complex threefold base. These 7-branes wrapped on internal four-cycles produce, at low energies, gauge group factors, and the specific intersections of these branes with suitable fluxes switched on leads to the matter content of the 4D model. Further intersections between these matter curves yields  Yukawa couplings. These ingredients result in Standard Model-like vacua, as well as many candidate extra sectors. Calabi-Yau fourfolds are a specific class of manifolds with a reduced metric holonomy group given by $SU(4)$; this means that the background has non-trivial Killing spinors. In the uncompactified directions these Killing spinor solutions are interpreted as preserved supersymmetries. In the dual M-theory description obtained by compactifying this 4D theory on a circle, one arrives at M-theory on 3D Minkowski space with an internal Calabi-Yau fourfold; in this dual description the inverse size of the compactification circle tracks with the volume of the elliptic fiber. The resulting 3D theory retains four real supercharges, i.e., 3D $\mathcal{N} = 2$ supersymmetry. 

The minimal case of 3D $\mathcal{N} = 1$ supersymmetry would instead result from M-theory compactified on a manifold with $Spin(7)$ metric holonomy. Explicit examples of such eight-manifolds include those given in \cite{JoyceSpin7}. These are obtained starting from a $T^8$ and taking a suitable quotient by a group action, i.e., an orbifold space. Locally, then, such quotients appear as elliptically fibered spaces, and so there is no issue with discussing such elliptically fibered non-compact $Spin(7)$ space. Globally, however, there cannot be such an elliptic fibration because such backgrounds do not correspond to a background which retains 4D $\mathcal{N} = 1$ supersymmetry. Rather, one at best has the weaker condition of a torus fibered eight-manifold which must degenerate along specific loci in the internal directions. Reference \cite{Heckman:2018mxl} argued that these degeneration loci can be interpreted in terms of five-branes wrapped on metastable five-cycles; i.e., they are wrapped on five-cycles which are topologically trivial in the full internal six dimensions, but which are nevertheless, locally volume minimizing configurations. In the 4D space-time these wrapped branes appear as point particles. These are the ``pixels'' of the construction.

Now, one of the primary observations in \cite{Heckman:2018mxl} is that there are in, fact, space-times in which 4D supersymmetry can be retained, albeit of a more exotic type. To see why, observe that in 4D Kleinian space-time $\mathbb{R}^{2,2}$, one can simultaneously impose a Majorana and a Weyl condition on spinors. As such, the minimal representation for spinors involves two real degrees of freedom. Obtaining a sensible supersymmetry algebra can also be arranged, but this turns out to require a partial breaking of the Lorentz group of $\mathbb{R}^{2,2}$. For cosmological scenarios, the natural candidate involves, in the original Lorentzian signature spacetime, either introducing a time dependent axion, or equivalently, a spatial three-form flux. This sort of ingredient is precisely what is also required to make sense of the non-holomorphic nature of the torus fibration present in the $Spin(7)$ backgrounds, and in particular the presence of wrapped five-branes (i.e., pixels).

In the limit of a large number of such five-branes, it is more appropriate to work in terms of the backreacted space-time where these branes have dissolved into flux. One then reaches a 4D Lorentzian signature geometry of the form $\mathbb{R}_{\mathrm{time}} \times S^3$, i.e., an Einstein static Universe. This configuration is unstable against perturbations, and inevitably leads to either a collapsing or expanding cosmology.

Let us now turn to the consequences of this scenario for observation. Perhaps the most basic feature of this pixelated scenario is that at high values of $\ell$, there is effectively a cutoff in the angular distribution of the power spectrum for scalar and tensor fluctuations. This is out of reach of current experiments (since $N$ is so large), but one could in principle envision a future experiment probing this feature. Of more direct relevance for current observations is the time dependence of $N$, the number of five-branes wrapped on internal cycles in the first place.

Since the original five-branes are wrapped on cycles which are topologically trivial in the internal direction, they specify metastable particles / pixels. The precise rate of winding / unwinding depends on how the winding / unwinding of the five-brane takes place in the six internal directions. In the limit where adding or subtracting a single pixel does not change the backreacted geometry very much, the growth rate / decay rate for pixels is expected to be roughly constant. On the other hand, in situations where modifying the number of pixels leads to more pronounced changes in the internal volumes of metastable five-cycles, one concludes that the pixel growth rate need not be constant. While it is therefore natural to expect string constructions to favor constant $\Gamma = \dot{N} / N$, the analysis of the present paper indicates that we obtain a better fit to the observational data by allowing the internal geometry itself to experience a different amount of backreaction after each pixel appears. This in turn suggests further deviations from a simple $Spin(7)$ background.


\bibliographystyle{apsrev4-1}
\bibliography{main}

\end{document}